Title: PMIPv6 Integrated with MIH for Flow Mobility Management: a Real Testbed with Simultaneous Multi-Access in Heterogeneous Mobile Networks

Affiliation:

- Hugo Alves, Instituto Universitário de Lisboa (ISCTE-IUL), Instituto de Telecomunicações, Av. Forças Armadas, 1649-026 Lisboa, Portugal;

- Luis Silva, Instituto Universitário de Lisboa (ISCTE-IUL), Instituto de Telecomunicações, Av. Forças Armadas, 1649-026 Lisboa, Portugal;

- Rui Neto Marinheiro, rui.marinheiro@iscte.pt, Instituto Universitário de Lisboa (ISCTE-IUL), Instituto de Telecomunicações, Av. Forças Armadas, 1649-026 Lisboa, Portugal;

- Jose Moura, jose.moura@iscte.pt, Instituto Universitário de Lisboa (ISCTE-IUL), Instituto de Telecomunicações, Av. Forças Armadas, 1649-026 Lisboa, Portugal.

ABSTRACT

The exponential growth of the number of multihomed mobile devices is changing the way how we can connect to the Internet. Our mobile devices are demanding for more network resources, in terms of traffic volume and QoS requirements. Unfortunately, it is very hard to a multihomed device to be simultaneously connected to the network through multiple links.

The current work enhances the network access of multihomed devices agnostically to the deployed access technologies. This enhancement is achieved by using simultaneously all of the mobile devices interfaces, and by routing each individual data flow through the most convenient access technology. The proposed solution is only deployed at the network side and it extends Proxy Mobile IPv6 with flow mobility in a completely transparent way to mobile nodes. In fact, it gives particular attention to the handover mechanisms, by improving the detection and attachment of nodes in the network, with the inclusion of the IEEE 802.21 standard in the solution. This provides the necessary implementation and integration details to extend a network topology with femtocell devices. Each femtocell is equipped with various network interfaces supporting a diverse set of access technologies. There is also a decision entity that manages individually each data flow according to its QoS / QoE requisites.

The proposed solution has been developed and extensively tested with a real prototype. Evaluation results evidence that the overhead for using the solution is negligible as compared to the offered advantages such as: the support of flow mobility, the fulfil of VoIP functional requisites, the session continuity in spite of flows mobility, its low overhead, its high scalability, and the complete transparency of the proposed solution to the user terminals.

I. INTRODUCTION

The number of mobile connected devices already exceeds the world population. Mobile traffic is increasing very fast and by 2019 three-quarters of mobile traffic will be generated by smartphones [1]. To support the growth of the traffic consumed by Mobile Nodes (MNs), new technologies, like LTE-A, are being deployed. LTE-A allows MNs to transmit up to 300 Mbit/s, but this is very dependent on congestion; its deployment is also very expensive and complex. To reduce the cost of using more spectrum or deploying more equipment in cellular networks, it is attractive for the operator´s infrastructure to also offer the available connectivity resources of already less expensive deployed infrastructures like WiFi, among others [2]. The coverage of these alternative wireless technologies is also growing. As a relevant example, the number of public WiFi hotspots is growing steadily fast [3]. Considering the previous aspects of cost and deployment's trend, the operators have a very strong motivation to divert mobile traffic from the core network to edge networks. In fact, it is foreseen that very soon more than half of the traffic generated each month by mobile devices will be offloaded from mobile networks by means of WiFi and femtocells [1].

Despite the possibilities already offered by existing technologies, operators aren't yet using all of the available network capacity to alleviate the load in the mobile infrastructure. This could be better achieved if operators could offer an enhanced connectivity service to individual mobile flows, with very distinct functional requisites and, independently of the used access technology. In addition, the end-users terminals have multiple network interfaces. Considering these novel aspects, the operators have to deal not only with more connected devices but also with other pertinent aspects such as, a large amount of data traffic, multiple network interfaces for each device, diversity of access technologies, and distinct Quality of Service (QoS) requirements depending on the used applications.

All these new requisites demand a new paradigm to route the traffic, and this challenge can be efficiently and effectively addressed by adopting the flow mobility concept [4]. The flow mobility concept is basically applying the mobility concept for each individual traffic flow i.e., each traffic flow can be routed individually through one of all the available routes. It also means the possibility of moving a specific IP flow from one MN access link to another. To support flow mobility the MN must be multihomed. i.e., it is connected to the network through multiple interfaces. In this way, the MN can exchange traffic with the network infrastructure simultaneously through all of its interfaces. Each flow can be routed according to some policy defined by the operator with or without the user intervention.

With a flow based mobility strategy, the operator can positively discriminate more critical traffic with the QoS available in the 3G/LTE-A networks when necessary, and simultaneously provides cheaper rates for non-critical traffic. For example, in a congested 3G/LTE-A cell an operator may only accept voice traffic and offload non-critical traffic, like HTTP or e-Mail, to the WiFi infrastructure. This strategy brings advantages to both operators and clients. In this way, the operators can solve the congestion problem imposed by the high surge of data traffic with hopefully an inexpensive solution and, the clients are offered with a set of services with enhanced QoS/QoE.

The flow mobility is also very relevant in networks that deploy femtocells, or sometimes called small cells, to extend service coverage indoors or at the cell edge, especially where network access would be otherwise limited or unavailable. The work presented here contributes with enhanced flow-mobility functionality by proposing and testing the simultaneous use of multi-access technologies on the same femtocell, addressing the feasibility and performance of having also each MN simultaneously connected through multiple technologies to the same or different femtocells.

Our current contribution extends the network based mobility approach and covers some novel aspects namely as follows:

− We have designed, implemented, and tested a complete solution operating in a real testbed to support mobility but assuming a flow granularity. We have also diminished the overload imposed by PMIPv6 over the network (see the final text in Section III. B);

− We have compared the performance of two distinct methods to identify new flows from the received packets, IPTables vs. NFTables (see IV. A). From our comparison results, we have selected IPTables;

− We have evaluated in a comprehensive way the performance of our proposal as well as its network overhead (see IV. B − D). We have evidenced that our proposal supports the requirements of traffic with strict QoS / QoE requirements. Our solution offers a negligible overhead at both the network (due to additional signalling traffic) and LMA / MAG nodes (due to extra processing load);

− Our solution fully supports the flow mobility (see IV. E). We have also evidenced that our proposal fully supports the requirements of VoIP traffic;

− Our proposal design already supports a decision entity that can analyse data flows to take routing decisions through diverse access technologies based on QoS/QoE flow parameters, user requirements and operator policies.

The next section, Section II, summarizes some of the state of the art of already existing mobile solutions; it describes the most relevant ones and make a comparison between them and the solution currently proposed. Section III explains how the proposed solution was designed and implemented, including the more relevant technologies. To validate the proposed solution, chapter IV discusses in a comprehensive way the obtained results from a network prototype with off-the-shelf equipment. Finally, chapter V summarizes and concludes the current work.

II. RELATED WORK

The mobility support is a pertinent requisite that needs to be urgently addressed in the current and upcoming networks. The deployed solution to support the mobility has a very important and different impact on the diverse network players, such as: content or service providers, network operators, users, and terminals [31]. In addition, each OSI protocol layer offers its own set of solutions to support mobile terminals. At the edge of the current Internet, the network layer is the most typical solution to route in a hop-by-hop way the packets among hosts. Consequently, we have selected for our work a Layer 3 mobility protocol.

The mobility support can be classified in host based or network based [33, 34]. These two approaches mainly differ in the responsibility of the MN tracking. Both of these depend on a central anchor point, the home agent. The home agent is located at the MN home network and routes the traffic directed to the MN, even if the MN isn't at the home network. The methodology of how the home agent knows the MNs location is what distinguishes both mobility solutions. In the host based, the MN must keep the home agent informed about its location in the network. On the other hand, in the network based solutions the MN doesn't need to keep the home agent informed about its location because there are entities at the network edge that can track the MNs movements and their locations and, maintain updated the home agent. With any one of these two referred solutions the MN can move across the network and the home agent will continue to route the traffic destined to the home network IP address to the foreign network IP address.

There are also distinct ways to support flow mobility [33]; there are solutions that in part permit the mobility of users flows either supported by the host [7], [8], [13] or by the network [9], [10], [16]. There is also one network based solution [11], [12] that proposes the concept of packet mobility and the introduction of network coding. This solution keeps the routing rules optimized in real time, by constantly collecting context information like traffic status, wireless channel characteristics, and the set of already transmitted data frames.

The Hercules stack [7], [8] is a mobility solution where a new network stack is deployed on every MN. This solution is suitable for devices that have multiple network interfaces. It has been developed a prototype for both Linux and Android platforms. Hercules Stack proposal consists in refactoring the network stack of all MNs that are involved in a communication, and it manages physical interfaces and provides a single virtual interface to the protocols above the IP layer. This solution creates an abstraction layer that hides the complexity of network management to the applications running on the MN. The management of the flows is made by a control plane that was added to the network stack. In practice, the operation mode of the Hercules works very similarly to Network Address Translation (NAT). It translates the private IP that it is used by the MN to an external IP that it is sent to the wire. One of the inconveniences of the Hercules is that it requires a new and complex network stack. It implies a very expensive operation to deploy it for all the MNs of an operator network. In addition, the translation mechanism inherits the classic network problems that are usually related to the NATs, such as the NAT traversal [30].

PMIPv6 [13 - 16] is a network based mobility protocol developed by IETF. Since it is a network based mobility protocol, it does not require the participation of the MNs in the mobility process. The PMIPv6 reuses most of the concepts defined in MIPv6 [17], the IETF standard to host based mobility. A PMIPv6 Localized Mobility Domain (LMD) is assured by two entities. The Local Mobility Anchor (LMA) is the entity located in the network core and it is responsible for anchoring the MNs locations in the LMD. The Mobile Access Gateway (MAG) is deployed at the network edge and it must track the MNs movements and keep the LMA updated about them. However, in its standard form, PMIPv6 does not support flow mobility. To overcome this limitation, the NETEXT Working Group is working on a draft proposal to extend PMIPv6 [18]. This draft proposal has been partially tested [9, 10] and promissory results suggest that PMIPv6 might be conveniently extended to support network based flow mobility. But many integration details are left out of the scope of the last referred internet-draft, namely the detection and attachment of MNs.

There is also a solution [11, 12] that uses the flow mobility concept by applying it to each individual packet and by using network coding to optimize the network performance. This solution adds a new entity to the MN, the User Information Server (UIS), responsible to collect network metrics that are later used to perform the packets routing. But the main novelty introduced in the work is that the routing decision can be taken for each packet individually. Despite of the relevant results shown in the work, it has not been proved that the added complexity scales for a full network based solution. In addition, the possibility of having out of order packets will either negatively impact the TCP congestion control or introduce reordering delay. All these induced effects can deteriorate QoE for users.

From the analysed solutions only some of them already support flow mobility [7, 10 - 12], but with some limitations. Our current work proposes to overcome some of these limitations or problems, as summarized in Table 1. We consider that requiring changes in the MN [7], [10], [12] is not a good strategy; the operator will also become dependent on the MNs and this hurdles an efficient management of flow mobility [32]. The complexity of implementing mobility at packet-level [11] might be overwhelming for the operator, and a better compromise may be achieved with flow mobility, as suggested by our current proposal. Additionally, a recent work discusses various mobility management schemes that propose the integration of the network-based PMIPv6 with IEEE 802.21 MIH framework for vertical handovers [32]. They also provide a systematic comparison of related efforts in the literature to evaluate and compare them, using some performance network metrics. Unfortunately, only one of the referenced proposals has performed tests with a real implementation scenario [23], and the remaining of them just has results obtained by simulation or theoretical analysis.

The work presented in this paper enriches the current literature with a more complete and solid integration of flow mobility in PMIPv6 [18], namely by improving the detection and attachment of mobile flows at the network edge by using IEEE 802.21 [6]. Contrary to previous work [32], the current contribution provides clear implementation details and extensive performance tests for femtocells scenarios using a real testbed, as it is discussed in the next sections.

### III. PROPOSED SOLUTION

The current paper extends the PMIPv6 protocol to support flow mobility transparency to users' terminals [32] in a novel scenario where all the terminals' interfaces are exchanging in parallel traffic with the Internet. The proposed solution also takes in consideration that it cannot require modifications in the MNs, it must maintain sessions active after handovers and it must have a performance very similar to the networking scenario with no flow mobility. To fulfil these goals, our work proposes a solution that extends PMIPv6 protocol, following the documents published by the NETEXT WG1. This group has already began the process of standardization for flow mobility extensions to PMIPv6 [18]. Figure 1 shows a typical scenario of how this proposal can be deployed. Here the Localized Mobility Domain (LMD) has one Local Mobility Anchor (LMA) and multiple Mobile Access Gateways (MAGs). The solution proposed is meant to be deployed both at the network core and edge.

The LMA unit is placed at the network core and it acts as the anchor point, i.e. Home Agent (HA), for the clients. The LMA has the conventional functionalities of mobility management and the necessary extensions to also support flow mobility. In this proposal the LMA is able to forward individually each client flow. The network core also includes a decision entity (e.g. Broker) that manages routes that the LMA applies for the flows. The decision entity may be either setup in proactive or passive mode. When it is passive, it just reacts to events in the network and makes handover decisions on the necessary flows. On the other hand, when it's in proactive it will constantly monitor the network by using QoS network sensors, to make timely handover decisions. This strategy can be useful when a flow needs to have guaranteed QoS during a specific time interval.

The MAG units are located at the network edge and are responsible for tracking the MNs movements. This proposal is also compatible with scenarios where single network equipment can provide multiple networks, e.g. femtocells that have both WiFi and 3G/LTE-A. In this case the client can be simultaneously connected through different access technologies, as shown in Figure 1. Here, a MN is connected with Correspondent Node (CN)1 and CN2 using different paths. The MAGs routes to the LMA are ensured by IPv6 tunnels. Since the focus of this work is to extend PMIPv6 to support flow mobility, we have decided to extend PMIPv6 open source software. This work has implemented the flow mobility extensions available from the Eurecom OAI-PMIPv6 software2.

---

[1] http://datatracker.ietf.org/wg/netext/documents/

There are two PMIPv6 implementations available as open-source software: the Eurecom OAI-PMIPv6, and OPMIP3. There is also a commercial version of the PMIPv6 protocol from Sibridge Technologies. The last solution has not been considered for this work due to the lack of public documentation and available results. Both OAI-PMIPv6 and OPMIP3 implementation were initially good candidates to receive flow mobility extensions proposed in this work. After analysing the compliance of the solutions with the PMIPv6 standard, performing some tests and code review, we have decided that the Eurecom OAI-PMIPv6 was the more suitable option to receive the flow mobility extensions. This option is also largely adopted by the main Linux distributions. Nevertheless, the OAI-PMIPv6 has a drawback in its method to detect the MNs attachments/detachments. It uses syslog messages that are generated by the access points. The problem here is that this approach is very dependent on the access technology being used. Different equipment may generate messages with quite different syntax on attachment/detachment events, or they may even not generate messages at all. To decouple this proprietary client management from the MAG, the current work has integrated PMIPv6 with an IEEE 802.21 protocol implementation. IEEE 802.21 is a standard that defines methods and technologies to achieve seamless handovers in heterogeneous networks. The integration with IEEE 802.21 makes our solution more robust and technology independent. IEEE 802.21 also improves the mechanisms for MNs movement detection, a key aspect for a good network based mobility solution.

The following sections describe in more detail the design/implementation of each relevant entity of our mobility solution, as follows: mobile nodes, local mobility anchor, and mobile access gateway.

### A. *Mobile Nodes*

There are multiple types of MNs, and this work will support some of them. When multihomed equipment configures its interfaces some problems may appear. After configuring all the interfaces, the node will have two sets of parameters; these parameters can be of two types. The ones that are bounded to the interface, like IP address and link layer address, and global parameters related with Domain Name Servers (DNS) and routing gateways. The main problem here is how the node will merge and use all the parameters available from the different interfaces. In this scenario a node may behave like a weak host, a strong host or it may use a LIF [19, 20].

Our work supports both the weak hosts and the nodes that use a LIF. The Strong host model is not supported because the host operating system using this model drops traffic if the ingress interface does not have configured the destination IP address present in the received packet. This impairs the correct functionality of our multihoming mobile scenario where all the host interfaces are exchanging data traffic in a simultaneous way with the Internet. Therefore, in spite of the strong host model offering higher security than the weak host model against a multihoming-based network attack, we have decided to not use the former in detriment of the latter. In addition, the security aspect is out of scope of our current work.

When the MN connects to the network, it is necessary to assign to it a home network prefix. This work assigns a unique prefix to each network interface. If each MN interface has a unique prefix or a set of prefixes, MAGs won't be able to route packets to the MN, if these packets are addressed to prefixes not managed by those MAGs. To avoid this problem, the LMA must explicitly inform MAGs with all the prefixes that are assigned to the MN, so that these MAGs can install routes to the MN.

---

[2] http://www.openairinterface.org/openairinterface-proxy-mobile-ipv6-oai-pmipv6
[3] http://helios.av.it.pt/projects/opmip

In the following sub-section, we will be discussing the local mobility anchor (LMA).

### B. *Local Mobility Anchor*

To support multiple flow bindings, the LMA Binding Cache (BC) was modified to map each individual route to a specific MN´s flow. In addition, the flow cache table was enhanced, to map each flow to an entry in the BC. All these changes turn it possible to define a specific route for each flow.

In order to speed up the forwarding for each individual flow, the LMA uses policy based routing. This routing provides a mechanism to make routing decisions based on all the information available from the packet header. This is achieved by using multiple routing tables and a forwarding label that is associated with a packet while traversing the network stack. This label is not part of the packet header; it only exists as a metadata in the LMA implementation kernel while the packet is in the system.

This packet labeling (marking) can be performed by any Packet Filter Framework (PFF) available in the LMA. At present both IPTables and NFTables from the netfilter project4 may be used, by setting a compile macro before LMA code compilation. With these PFFs the packet processing takes place in the kernel space, which gives a high efficiency. However the LMA works in the user space and packets have to be somehow diverted from the kernel to the User Space, as shown in Figure 2.

But then again, diverting all packets to the User Space is not efficient. To minimize the impact on the system, the PFF only diverts the first few packets of the flows that haven't been previously analysed, and places them in a User Space queue (i.e. PUQ in Figure 2). The packets placed here will not be routed until some user space application applies a decision to them. For example, when the first packet of the flow A enters into the system it will be received by the kernel. The kernel will try to find a rule to mark that packet. Since this is the first packet, and there isn't yet a rule configured to mark it, then this packet is sent to the User Space queue.

In the User Space, LMA will parse the packet to identify to which MN it's meant, shown in Figure 2 as the Flow Identifier Module (FIM). To manipulate packets that are placed in a User Space queue, the libnetfilter_queue5 has been used.

After identifying the MN, FIM sends the flow identification to the Flow Scheduler Module (FSM). With this information the FSM creates in the kernel a forwarding rule that marks all the packets of that flow with a forward mark. Finally, the Mobility Manager Module (MMM) is the module that is monitoring the network conditions and reacts to its changes by changing the installed flow routes. The MMM will also interact with the routing decision entity.
After this, when the next packets of flow A arrive, there is already a matching rule installed. The packet is not going to be managed anymore at the User Space and it will proceed with its normal path in the kernel. With this technic a better performance is accomplished for all the flow packets with the exception of the first one.

---

[4] http://netfilter.org/projects/nftables/
[5] http://www.netfilter.org/projects/libnetfilter_queue/

The internal modules of LMA developed by us, which are visualized in Figure 2, such as FSM, MMM, and FIM, are comprehensively discussed in the following sub-sections.

*Flow Scheduler Module (FSM)*

The FSM is a module that receives information about a flow and chooses a route for it. This entity communicates with a decision entity [22] [35] that can provide information about the network status and the MN traffic requirements. The proposed solution is prepared to work with any implementation of this entity. The only requirement is that the provided entity should use the FSM interface. For testing purposes, the FSM is also capable of choosing by its own decision one path either permanently or randomly from the set of available paths.

*Mobility Manager Module (MMM)*

The MMM is an event driven thread that waits for events related with the mobility process, like Proxy Binding Updates (PBUs) that are triggered by the MAG when a MN moves or attaches a new interface. This module is at the core of the PMIPv6 implementation.

Figure 3 shows the finite state machine of the MMM. Whenever the MMM receives a PBU message it will search its Binding Cache (BC) for a Binding Cache Entry (BCE) belonging to the node referenced in the PBU. The MMM identifies a BCE by the tuple MN ID<->Serving MAG address. Note that there is one BCE for each MN connected interface. In this approach it is not possible for a client to be connected to the same MAG by two or more distinct interfaces. For example, in a normal situation where a MN is using two interfaces, the MN should have two entries in the binding cache, one for each MN interface ID mapped with a distinct serving MAG. In the proposed architecture this is not a problem since a heterogeneous femtocell with multiple technologies may deploy one MAG per access technology.

After searching for the BCE, the LMA may perform one of four different actions, as detailed in Figure 3: register the attachment from a MN (1), renew the lifetime of a BCE (2), perform a handover (3) or delete a BCE (4). The attachment (1) deals with connections of MNs network interfaces. This event is triggered by a PBU message. If the PBU is valid, the LMA will create a new BCE in the binding cache structure and setup the necessary routes and tunnels to send the traffic to the MN. This action ends by sending a Proxy Binding Acknowledge (PBA) message to the MAG. To comply with the PMIPv6 standard, the BCE must have a lifetime to graciously discard old routes that are no longer active. The renewal (2) of a BCE is triggered by receiving a PBU with a lifetime with a positive value, when there is already a BCE referenced by that PBU. When receiving a PBU that references a different MAG for a MN interface ID, than the one present in the BCE, it means that the MN has changed its attachment and a handover must be performed. In this case (3) the LMA will update the old BCE and the routes associated with it. Both the renewal and the handover process ends by sending a PBA message to the new MAG. Finally, the delete action (4) is triggered by receiving a PBU with a lifetime value of 0; this means that the MAG is requiring the deletion of BCE associated with that MN interface ID. In this situation, the LMA will delete the BCE and respective routes. After this, a PBA message is sent back to the MAG that previously sent the lifetime message with the zero value.

Both handover (3) and delete (4) actions have impact on the flows that are destined to the MN. In these situations, the LMA will have to reschedule all the flows of the associated client. This is done by sending a request to the FSM to reroute the MN flows.

*Flow Identifier Module (FIM)*

The last component is the FIM that processes data packets addressed to MNs that belong to flows that have not yet an assigned routing rule. It extracts packets that are in the user space queue and identifies to which MN they are addressed. To parse a packet, the FIM uses a traffic selector, a 6-tuple selector that it's composed by: source and destination IPv6 addresses, source and destination transport protocol ports, transport protocol and flow label extracted from the IPV6 header. These values are enough to uniquely identify a flow. They are also extracted very efficiently since these are all readily available in the packet and segment headers. By querying the BC with the destination IPv6 address the FIM can identify to which MN the flow belongs. When the MN is identified, FIM passes the traffic selector and the MN identification to the FSM so that it can create a routing rule to that flow.

To inform the MAGs about changes in the mobility sessions, the LMA must normally send an Update Notification Message [21] to each MAG to inform it about new prefixes that the MAG should route. As a novelty, our proposal simplifies this process and it does not use the previous update message. In fact, since all the traffic addressed to the MNs is tunnelled from the LMA to the MAG, each MAG can simply forward all the traffic coming from the tunnel to the wireless medium. In this way the network overhead is conveniently reduced.

In the following sub-section, we will be discussing the mobile access gateway (MAG).

### C. *Mobile Access Gateways*

In this proposal the MAG implements the standard PMIPv6 functionalities; however PMIPv6 doesn't define how the MAG should detect MNs movements. For this reason, we propose the use of the IEEE 802.21 framework to decouple the attachment management from the PMIPv6 protocol. With this proposal the MAG will communicate with a Media Independent Handover Function (MIHF) deployed in the network. This communication is based on triggers, and to receive these triggers the MAG makes a registration in the MIHF for attachment and detachment events generated on the femtocell. These events are detected by Link Saps (LSs), installed for each access technology, which will forward them to the MIHF that then will forward them once again to the MAGs that have subscribed them. This work has enhanced the ODTONE [23] IEEE 802.21 implementation to achieve this.

Figure 4 shows the high level architecture of the MAG. This is composed by 5 modules: the Finite State Machine (FSM) (1), the events handler (EH) (2), the Binding cache (BC) (3), the MIHF (4), and the LSs (5). The top layers are FSM (1), EH (2) and BC (3) that participate directly in the PMIPv6 flow mobility. The FSM (1) is core to implement the enhanced PMIPv6 protocol; its main tasks are to process messages, applying protocol rules, and to react to events, like attachments and detachments. The EH (2) listens for events on the network: PBA messages, ICPMv6 messages and MIHF notifications. During the setup phase this module asks to the MIHF to be notified about Link up and Link down events that can occur in the network interfaces under the domain of this MAG. The BC (3) is a data structure that stores all the information necessary to reach the MNs, e.g. the most relevant are the Home Network Prefixes (HNP)s assigned to the MN. In this way, the BC of the MAG entity is different from the BC of the LMA entity.

The MIHF (4) is another IEEE 802.21 logical entity that acts as a central point to receive events from the Link Saps and forward those to other logical entities that query network conditions or subscribe notifications. In practice it acts as an abstraction layer to applications that want to manage heterogeneous networks.

The LSs (5) are abstract media dependent interfaces from the IEEE 802.21 standard, and as already referred these were developed for each access technology. As such, each LS interacts with one physical network interface to detect events that are occurring in the associated network. In this work the events are the link up and link down, i.e. attachments and detachments of MNs, but other information could also be used.

Figure 5 shows a simplified version of the MAG FSM (1). This shows the MAG logic used to process the events that are necessary to implement flow mobility. There are four events: attachment (1), detachment (2), PBA response to a register PBU (3) and a PBA response to a deregister PBU (4). They are respectively triggered when a client connects a new interface (1), when it disconnects the interface (2), and when the MAG receives a PBA, either related with a registration (3) or with a deregistration (4). The attachment (1) and detachment (2) events can be generated by multiple sources. As described above, the preferred mechanism for this is by using the IEEE 802.21 framework, but syslog messages at kernel level may also be used. Nonetheless, only one of these mechanisms can be active at a time.

When the MN is connected for the first time to the MAG an attachment event (1) is received. In this situation the MAG gets the MN Network Access Identifier (NAI) from an Authentication, Authorization and Accounting (AAA) server and creates a temporary Binding Cache Entry (BCE) for the MN. Next it sends a PBU register message to the LMA. It is expected that the previous sent PBU message will trigger a PBA in response. When a MAG receives a PBA, it tries to find a corresponding BCE for the MN. The search can either return or not a BCE. If there is a BCE mapping it can either be a temporary or a permanent entry.

The temporary entry exists while the MAG hasn't received the PBA message in response to a PBU. When a PBA signalling a success is received, it will change status of the entry from temporary to permanent and it will send a unicast router advertisement containing the HNP to the MN. This concludes the registration process. If the PBA is received with an error status code, the BCE entry is deleted. If the MAG finds a permanent BCE it means that the PBA is the response to a previous renewal request and as such the BCE lifetime will be updated.
At last, the MAG may also receive a detachment event. This situation occurs when the MN disconnects from the network. In response to this event the MAG deletes the route and the BCE that references the MN. It then sends to LMA a deregister PBU to notify the MN detachment.

The MAG also manages the lifetime of each BCE, and before its lifetime expiration the MAG must check if the MN is still reachable. This is done by sending a neighbour solicitation message to the MN. If the MN is still active, i.e. it responds back with a neighbour advertisement, the MAG will renew the BCE. Lifetime maybe a dynamic value, that is calculated based on the volatility of the MN. If the MN is a mobile node with a significant speed, the lifetime should have a smaller value. Conversely, less volatile (static) MNs should have a higher lifetime value to avoid unnecessary signalization messages. Our choice of the neighbour discovery method is supported by a recent analytic study [32] where it is shown this method offers lower latency and fewer losses than other methods to discover the network topology and enable the network selection.

The two MNs supported types, weak host and LIF, have some particularities during HNP assignment. In a flow mobility scenario, the MN must have multiple prefixes for its interfaces. NETEXT draft [18] specifies three use cases for the prefixes attribution. The MN either always receive the same prefix when it attaches an interface, or it can receive a new prefix for each attachment. The third option is a combination of the last two, where a new or a previously assigned prefix is assigned, depending on network policies. The proposal presented here adopts the second use case, where the MN receives a unique prefix for each interface. If each MN interface has a unique prefix or set of prefixes, MAGs won't be able to route packets to the MN, if they are addressed to a prefix that they don't manage. In this case, a MAG cannot forward packets addressed to a prefix that it doesn't know. In this situation the LMA must explicitly inform MAGs with all the prefixes that are assigned to the MN, so that they can install routes to the MN.

In our proposed solution, we have enhanced PMIPv6 with a standard abstract layer (L2.5) that can receive events from diverse access technologies (L2) reporting technology status to the MAG entity. The MAG empowered with this status information per technology can perform a more efficient mobility flow management among all the available technologies. In this way, the next sub-section describes in a more detail way how we have implemented IEEE 802.21 inside the MAG to support flow mobility in multihoming scenarios.

*Integration with IEEE 802.21*
This sub-section discusses how IEEE 802.21 works inside the MAG to enhance the operation of PMIPv6. In this way, Figure 6 details how 802.21 can be used to detect MNs movements in the network to deploy flow mobility.

Before receiving any network generated events, the MAG must register with the MIHF. This implies the exchange of several messages. The first message (1) registers the MAG as a client in the MIHF which triggers an optional acknowledgment response (2). After a successful registration, the MAG sends a capability discover request message (3) to the MIHF. The MIHF sends back a capability discover response with the interfaces that it manages and the available events for each one (4). Then the MAG checks if the interface that it should manage is under that MIHF supervision. If it is, it sends an event (link up and link down) subscribe request (5) to the MIHF. This message requests to the MIHF to send back a message when it detects an attachment or a detachment on that specific link. After receiving confirmations (7) (8), the MAG is waiting for event messages from the MIHF.

When a MN connects to an access managed by a Link Sap it will trigger a LinkUP attachment event message (9). This event is then sent to the MAG by the MIHF (10). When received, the MAG only identifies the link layer address of the MN that has been attached. To obtain the MN ID, the MAG must convert the MN link layer address to the EUI-64 format and send an authentication message to the AAA server (11). The AAA server then returns the MN authorization, the MN ID and the HNP that should be assigned to the MN (12).
To finalize the attachment, the MAG sends a PBU message (13) to the LMA with the information obtained from the AAA server. After receiving the PBA (14), the MAG advertises the prefixes assigned to the client by sending a unicast Router advertisement (15) message to the MN.

Integrating PMIPv6 with IEEE 802.21 improves the MNs movement detection mechanism. It also opens the door to future upgrades that can take advantage of all the functionalities that IEEE 802.21 has to offer. For example, in an IEEE 802.21 scenario the MIHF can be used to detect situations of an eminent link down. This information can be passed to the LMA so that it can start preparing in a proactive way handovers for flows that will be affected by the link down event.

## IV. EVALUATION RESULTS

The current proposal has been developed and a significant number of tests has been performed with real equipment. In the following text, some details are provided regarding tested software, used equipment, configurations and the multiple tools used to perform the tests and process the obtained results.

The testing topology consisted in a network infrastructure with one LMA and four MAGs, one client and one correspondent node, as shown in Figure 1. The four MAGs are running in two physical nodes. In this setup, MAGs were deployed in embedded devices that usually work as home wireless gateways, i.e. TP-LINK WDR-4300 routers, and they were installed with a customized version of the OpenWrt operating system. This type of routers is relevant to speed up prototyping and testing because of their capabilities like: low cost, extensive documentation, open source software and the support for plug-and-play external hardware (e.g. USB dongles). They also make a good use case for residential gateways or other access routers that may work in a femtocell configuration. For instance, having a Linux operating system installed has permitted the configuration in a single equipment of two independent APs with different technologies, enabling a multi technology femtocell. The routers have been installed not only with the MAG software but also with one MIHF and two Link Saps. The TP-LINK WDR-4300 provides two independent WiFi networks, one at 2.4 GHz and another at 5GHz band. This feature was very useful to evaluate the multi-technology femtocell that is proposed in our work.

In the performed tests, the client is a regular laptop that has been configured with two WiFi USB dongles. The client, the correspondent node and the LMA were off-the-shelf computers installed with the CentOS operating system. More details about the hardware are detailed in Table 2.

Several well-known tools have been used during the tests. For example, the tool IPERF[6] has been used to generate the network load. However, to measure the time to forward a packet a custom measurement toolset has been developed. The packet forward time must be measured in the most accurate way as possible. Ideally, measurements should be made by external hardware/software that doesn't interfere with the equipment running the PMIPv6 software. To minimize this problem, a kernel module was developed to extract the necessary information from the packets that are routed by the LMA and the MAG. With a kernel module it's possible to obtain results with an adequate precision without a significant performance impact on the equipment running the module. The developed software uses the netfilter framework present in the Linux Kernel.

Our evaluation tests are summarized in Table 3 together with their main goals. The obtained results from each test are individually discussed in the following sub-sections.

### A. *Choose the more convenient method to filter packets*

As previously explained, the proposed solution may use one of two packet filtering frameworks, IPtables or NFtables. Results comparing both frameworks are then necessary. In particular, it is necessary to collect data to evaluate these two frameworks regarding packet forward time and the time taken to create a rule associated to a new flow after other flows are already present within the system.

---
[6] https://iperf.fr/

The packet processing time was measured on a single data flow, from the CN to the MN, for an increasing number of forward rules that are already present in the routing tables and for that reason the kernel is required to process this. The number of forward rules increases at regular intervals and for each increase, 100 samples were taken. The flow had a constant bit rate at 100 kbps.

As shown in Figure 7, IPTables presents the best behaviour, without a noticeable delay. In an extreme scenario, with 1000 rules to process, IPTables can forward a packet in 75us and NFTables in approximately 360us. These delays are extremely low. Even when considering VoIP conversation where the maximum one-way latency for a high quality conversation should be at maximum 150 ms [24], [25], which is completely fulfilled by our system.

It is also important to validate that processing new packet-filtering rules, a task executed by the (LMA) FSM, is fast enough. Figure 8 shows the time that it takes FSM to setup the rule n+1 when n rules are active. For NFTables the time to insert a new rule does not increase with the number of installed rules. In fact, this value remains constant at a high delay of 180ms. IPTables insertion time for new rules starts with a much lower value of ~25ms and increases up to ~75ms when 1000 rules are present. These processing delays are not so much negligible; however they might only impact on a few packets to be forwarded, i.e. typically the initial packets of each flow are the only ones affected by this aspect. With these results, IPTables has been proved to be the best option for forwarding flow packets, even when there is a considerable amount of rules installed in the system.

## B. *Study the processing overhead associated to the initial packets of a new mobile flow*

The solution proposed in this work requires that the first few packets of a flow pass through a User Space queue before being routed to the destination. It is then necessary to know the penalty of diverting these packets to the user space for their classification. A test to measure this impact is necessary. In particular, this test has to measure the packet forward time of the first few packets of a flow, where a higher packet forward time is expected. One should notice that packet forward time will decrease significantly after the User Space software has installed a routing rule in the kernel. The reason for this behaviour seems obvious: a kernel module runs faster than a corresponding module in the user space.

Results were then obtained for an increasing number of flows. The test started with one flow at 100 Kbps, and regularly 100 kbps flows were added, until 50 flows were present. Figure 9 shows the measurements made for the first set of n packets of each new added flow. It shows that the time to process packets diminishes rapidly after the first processed packet of the new flow. After the fifth packet the packet process time is less than 40 µs. This test has shown that there is an extra cost to process the first packet of a new flow, as it was expected. But that cost starts to get rapidly diluted after the 5th packet of a new flow. Figure 9 also proves that packet forward time does not seem to increase with the load of the system, e.g., processing the first packet of the 50th flow takes approximately the same time as processing the first packet of the first flow. It is important to notice that when there are 50 flows, the aggregate throughput in the LMA is around 5 Mbps.

Taking into account all of the advantages of the flow mobility, the delay increase of the first set of packets is tolerable. This cost can be reduced by having more capable machines or by tuning the packet classification framework in Linux.

## C. *Evaluate the performance of MAG and LMA entities deployed in a real testbed*

This test focused on measuring the delay introduced by the LMA and the MAG in the network. The results shown in Figure 10 were measured in a stable situation where all the flows were already processed by the LMA user space software. In this situation the only entities involved in the packets forwarding is the kernel and IPTables.

Figure 10 shows the cumulative probability distribution function results for the packet processing times of the MAG and LMA for different throughputs. Figure 10 shows that packet forward time in the MAG is not noticeable affected by the throughput at least until 100 Kbps per flow. For example, with 50 flows at 100 Kbps, i.e. an aggregated throughput of 5 Mbps, the MAG can forward 90% of the packets in less than 15 µs. This was the expected behaviour for the MAG, since in this scenario the MAG is just acting as a bridge between the LMA and the MNs, and it doesn't make any significant task that could delay the packets.
On the other hand, the LMA presents more distinct results. The first thing to notice is that the forwarding times of the LMA are approximately twice the ones got on the MAG, mainly for low traffic scenarios. The reason to this behaviour is the complexity of the LMA that has a set of forwarding rules, and logic that it implements, that have to be applied and make it more complex than the MAG. For example, the LMA must check all the incoming packets to verify if the flow has already a defined route.

For the sake of the clarity of the results, another relevant aspect to notice in Figure 10 is that the higher throughputs have apparently lower packet process times. This could be a contradiction to the expected behaviour, since it was expected that packet processing would increase with the load of the LMA. In reality, the phenomenon that it's shown in Figure 10 it's an optimization made by the operating system to improve the performance in high throughput networks, as we following explain.

### *NAPI effect on LMA´s packet processing time*

Most operating systems use an interrupt based mechanism to react to hardware events. When the network interface receives a frame it places that frame in a buffer, which is shared with the operating system. After this, the network interface sends an interrupt to the CPU. The CPU detects the interrupt and immediately calls the correspondent driver that handles that specific interrupt, and the frame processing begins.

It's rather obvious that with high throughputs the interrupt rate may be excessively high. This is a problem because the hardware interrupts have priority over all the other tasks executed by the operating system. And when this happens the CPU must spend precious time saving the actual state so that, after the interrupt processing finishes, it can return to the previous status. In networks with a high packet arrival rate this behaviour will slow down the system [26], [27]. In a worst case scenario the system may drop throughput to 0, when the CPU is spending more time processing the interrupts than performing other tasks [28].This is a serious problem in high throughputs networks, e.g. gigabit networks. To overcome this problem the Linux has introduced a new mechanism to avoid the high number of interrupts, the New API (NAPI) [29]. This mechanism removes the inconveniences of having one interrupt for each packet.

With NAPI, the driver has to provide a buffer, either shared or not with the operating system, to store the frames received from the wire, normally a ring buffer. Additionally, it has to provide a poll method that retrieves received frames from the network interface. This method can be invoked by the networking subsystem of the Kernel when it wants to get a batch of frames to start the processing in the network stack. Basically the interrupt based strategy is replaced by a poll mechanism.

The NAPI behaviour is exposed in the Figure 10 plots. As an example, for a load of 50 flows and throughput of 10 Kbps, when the NAPI optimization is not relevant, over 90% of packets have a forwarding time in LMA slightly above 30 µs. This value is already quite low. But as soon as throughput starts to increase, NAPI optimization impact becomes more visible, and packets processing time converge to a range values around to 10 µs (i.e. throughput of 100 Kbps). In this way, as low throughput traffic is expected in some network nodes (LMAs or MAGs), then NAPI can be disabled in those nodes to slightly decrease the packet forward time. But doing so, node processors could become more overloaded.

### D. *System's overhead induced by our proposal that uses a personalized routing rule to manage each mobile flow*

It is important to measure the impact of the added complexity in this solution, when compared to the standard PMIPv6 protocol. Figure 11 compares performance when flow mobility is either enabled or disabled in the implemented solution. Multiple active flows in both scenarios were considered, and packet process time in the MAG and the LMA were measured. As expected, packet process time at the MAG is identical in both cases. This is justified because the MAG hasn't suffered major changes to enable flow mobility.

On the other hand, since the LMA implementation has major changes, a performance penalty was expected. Despite this, Figure 11 shows that the impact in performance is almost negligible. For example, with 5 flows at 100Kbps, around 90% of the packets are forwarded under 30µs, but without flow mobility implemented the alternative is only faster by around 5µs. Despite the extra delay, this test shows that the extra cost of enabling flow mobility is almost insignificant when compared to the advantages brought to the clients and to the mobile operator.

### E. *Handover delay of a mobile flow*

One of the features offered by the proposed solution is the capability of moving a specific flow from one access to another, when the MN is multihomed. The handover can either be triggered by QoS preferences or by a link down situation. Moving an ongoing flow from one access to another must be made with caution. The handover should take the least time possible so that the MN´s flow suffers a minimum loss of packets.

A test was setup with a MN with two network interfaces and each interface connected to a different MAG. One of the interfaces receives a specific number of multiple flows, at 100 Kbps each, and the other interface receives only one flow, also at 100 Kbps, which is moved after a network detachment event.

The handover time is measured as the time difference between the last packet received in the interface that went down and the first one received in the new interface. Obviously, both packets that are taken into account in the measurements belong to the same flow that undergoes the handover. This measure is not a real handover time, since it includes the packet inter-arrival time. Given that in the tests data flow has a constant bit rate, the period between packets is approximately constant. If this period is subtracted to the inter-arrival time measured during the handover, a reasonable accurate measure of the handover time is obtained. This measurement methodology provides better results for high throughputs when the period between consecutives packets is very small. For low throughputs the period between packets is relatively high, when compared to the handover time, and in this case, the measure would be less accurate.

Results for this test are shown in Figure 12. The maximum handover time is between 50 and 150 ms. These values cannot be neglected, but for some types of traffic and/or transport protocols they are perfectly tolerable, mainly if retransmissions are possible.

The more important results obtained after running our five evaluation scenarios are compiled in Table 4.

V. CONCLUSION

Our work successfully implemented in a prototype some mobility extensions for PMIPv6, including the integration of the IEEE 802.21 standard. The current proposal enriches the literature because according to our knowledge the vast majority of previous proposals have only provided simulations or analytical results [32] and none of these supported flow mobility [7, 11-12] in a complete transparent way to the terminals such as our proposal effectively does. In this way, our work greatly contributes with relevant and comprehensive performance results obtained through a real testbed, where our solution was completely deployed to manage mobile flows using personalized routing rules. Our results prove that the main initial goals of our proposal are completely fulfilled, including the track of mobile flows and the deployment of personalized routing rules for mobile flows. In addition, all this new functionality was deployed with a marginal negative impact on the system's real performance, suggesting that our implementation is perfectly scalable.

Our contribution also shows that it is feasible to deploy femtocells that provide simultaneous multiple accesses to multihomed mobile nodes. These femtocells can be implemented in low cost embedded devices that can be easily deployed as access routers at the network edge (e.g. residential or home scenarios). With these femtocells the operators can increase theirs network coverage at a low cost.

For future work, we propose to explore the integration of a decision entity [35] with the LMA entity. The developed prototype has already a basic flow scheduler but its behaviour can be further enhanced. As an example, the decision entity may collect information about network conditions and flows' requirements and optimize the operation of the flow scheduler. With this it is perfectly possible to optimize both the network operation and the QoE of flows.

ACKNOWLEDGMENT

Work supported by Fundação para a Ciência e Tecnologia (FCT) of Portugal within the project UID/EEA/50008/2013

*Table 1 – Novelty of our work*

| Problems / Limitations of previous work | Major contributions of our work |
|---|---|
| Most part of the previous work was host-based impairing an efficient deployment for a large number of systems with heterogeneous terminals | The current contribution is a network-based solution offering a complete abstraction to terminals; it hides from the hosts all the details in how the mobility should be controlled. Our solution is more scalable than previous contributions because the former manages flows. |
| The previous results were obtained mainly from simulation and analysis | Our results were obtained through a real testbed |
| Previous contributions assume the usage of a single technology during a specific time interval | Our work assumes the usage of multiple access technologies in a simultaneous way |
| Previous work supported terminal mobility | Our work supports flow mobility |
| Previous work assumes the same routing rule for all the flows of the same terminal; Other work processes individually each packet. | We assume a personalized (and more efficient) routing rule for each mobile flow |
| Previous work assume a reactive mobility support (of terminals) | We assume a proactive and more intelligent/efficient mobility management (of flows) |

*Table 2 - Testbed hardware*

| Hardware | CPU | RAM | Network | OS | Function |
|---|---|---|---|---|---|
| TP-Link WDR4300 | Atheros AR9344 560MHz | 128 MB | Gigabit Ethernet | OpenWrt 12.09 | MAG |
| CN | Intel P4@2.4GHz | 942 MB | Fast Ethernet | CentOs 6.5 | Correspondent node |
| LMA | Intel E5335@2GHz (x2) | 4GB | Gigabit Ethernet | CentOs 7 | LMA |
| Client/MN | Intel 3537 | 4GB | Gigabit Ethernet | LMDE | MN |
| Alfa AWUS036H | - | - | WiFi 2.4GHz | - | Client Interface |
| TP-Link TL-WDN3200 | - | - | WiFi 2.4/5GHz | - | Client Interface |

*Table 3 – Evaluation tests through a real testbed*

| Identifier | Designation | Main Goal |
|---|---|---|
| A | Choose the most convenient method to filter packets | Performance study (i.e. processing delay) between IPTables and NFTables to discover a new flow and create a new routing rule for that flow; the final goal is choose the best option. |
| B | Study the processing overhead associated to the initial packets of a new mobile flow | Processing time evolution along the initial packets of a flow that are managed by applying the same routing rule |

| Identifier | Designation | Main Result |
|---|---|---|
| C | Evaluate the performance of MAG and LMA entities deployed in a real testbed. | Study of packet processing time in both MAG and LMA units |
| D | System's overhead induced by our proposal that uses a personalized routing rule to manage each mobile flow | Studying the additional latency imposed by packet filtering and routing rules over mobile flows |
| E | Handover delay of a mobile flow | Study the performance of our solution, based on PMIPv6 enhanced with MIH, after a network failure to find a new flow route |

*Table 4 – Obtained main results from our testbed evaluation*

| Identifier | Designation | Main Result |
|---|---|---|
| A | Choose the most convenient method to filter packets | IPTables has proved as the best option to diminish the system delay that is associated to create and apply a specific routing rule to each flow (both @ LMA) |
| B | Study the processing overhead associated to the initial packets of a new mobile flow | The processing overhead is rapidly diluted after the 5th packet of a new flow |
| C | Evaluate the performance of MAG and LMA entities deployed in a real testbed. | Study of packet processing time in both MAG and LMA units; NAPI optimization impact on LMA forwarding time was noticeable only for high throughputs |
| D | System's overhead induced by our proposal that uses a personalized routing rule to manage each mobile flow | Negligible performance penalty @LMA induced by flow mobility controlled by our proposal that filters packets and then applies routing rules to filtered packets |
| E | Handover delay of a mobile flow | Handover time was within the range [50, 150] ms; this range is perfectly acceptable, as an example, for VoIP |

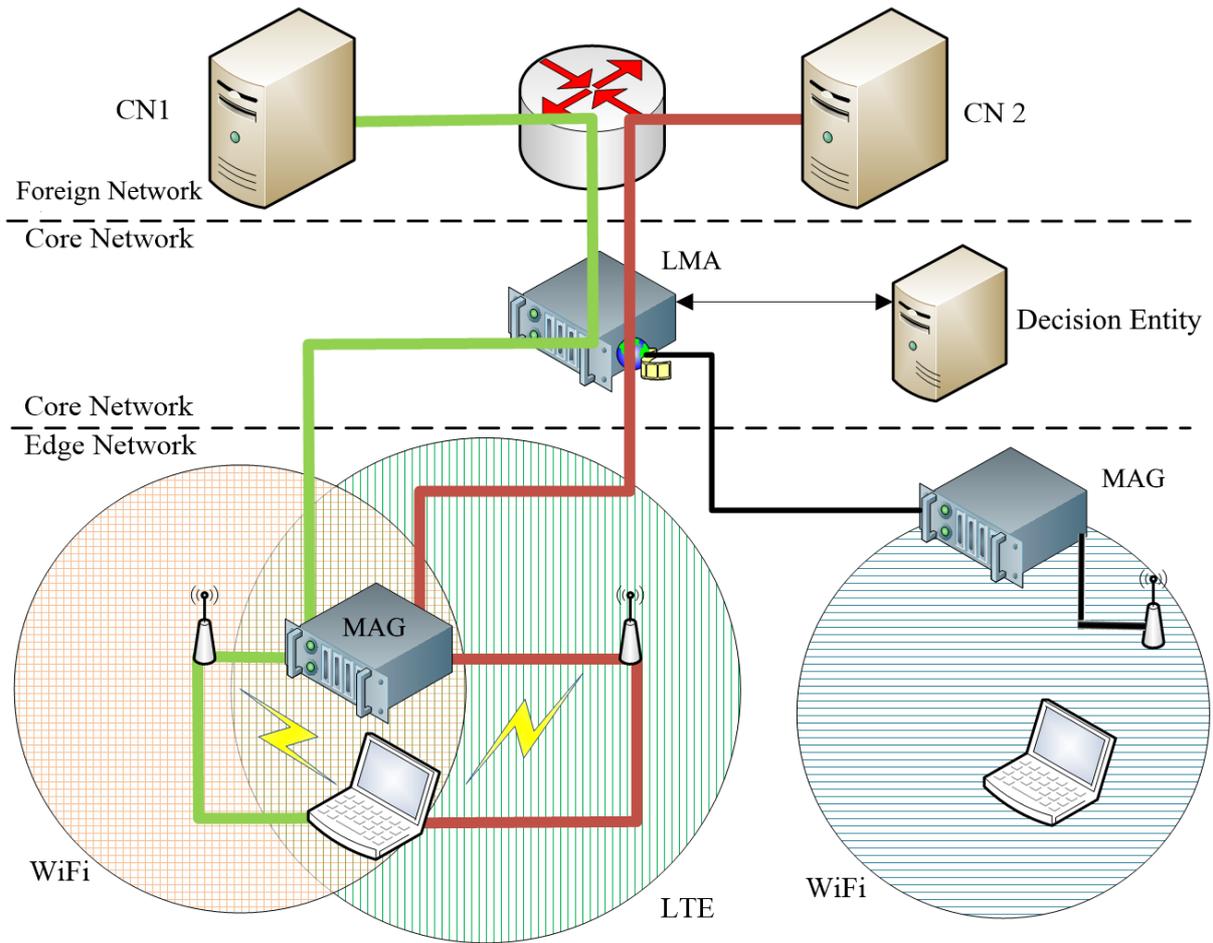

*Figure 1 - Typical deployment scenario of PMIPv6 in a scenario with mobile flows*

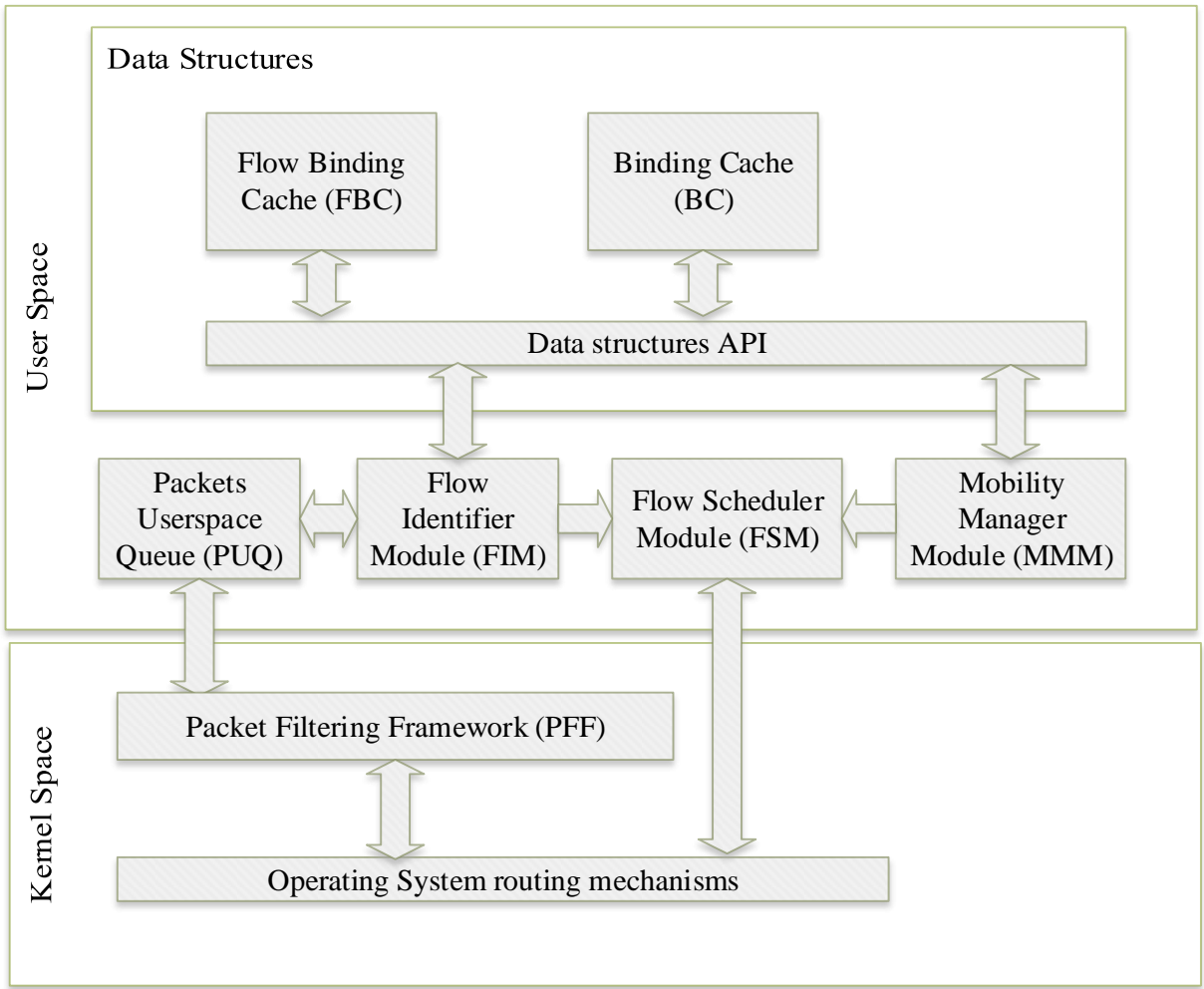

*Figure 2 - LMA block diagram*

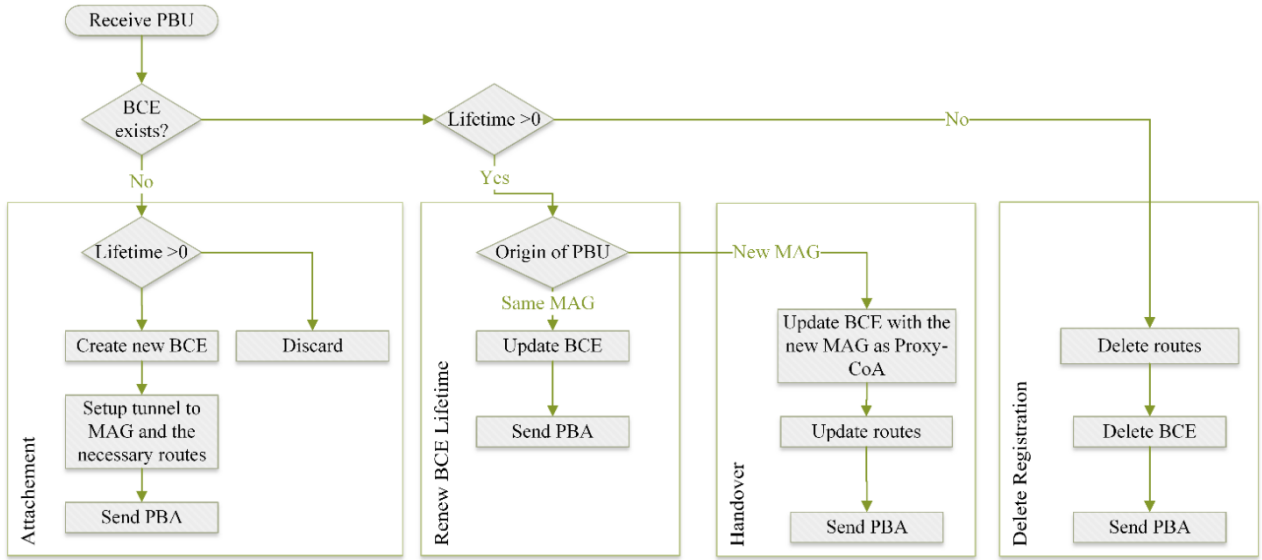

*Figure 3 – Finite State Machine of LMA's Mobility Manager Module*

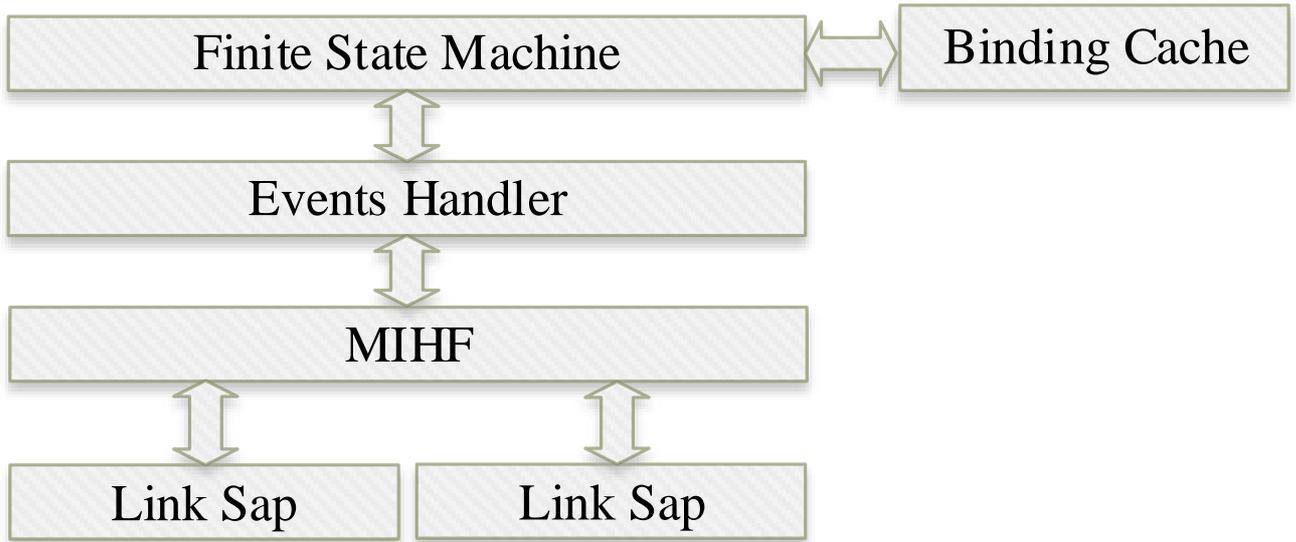

*Figure 4- MAG block diagram*

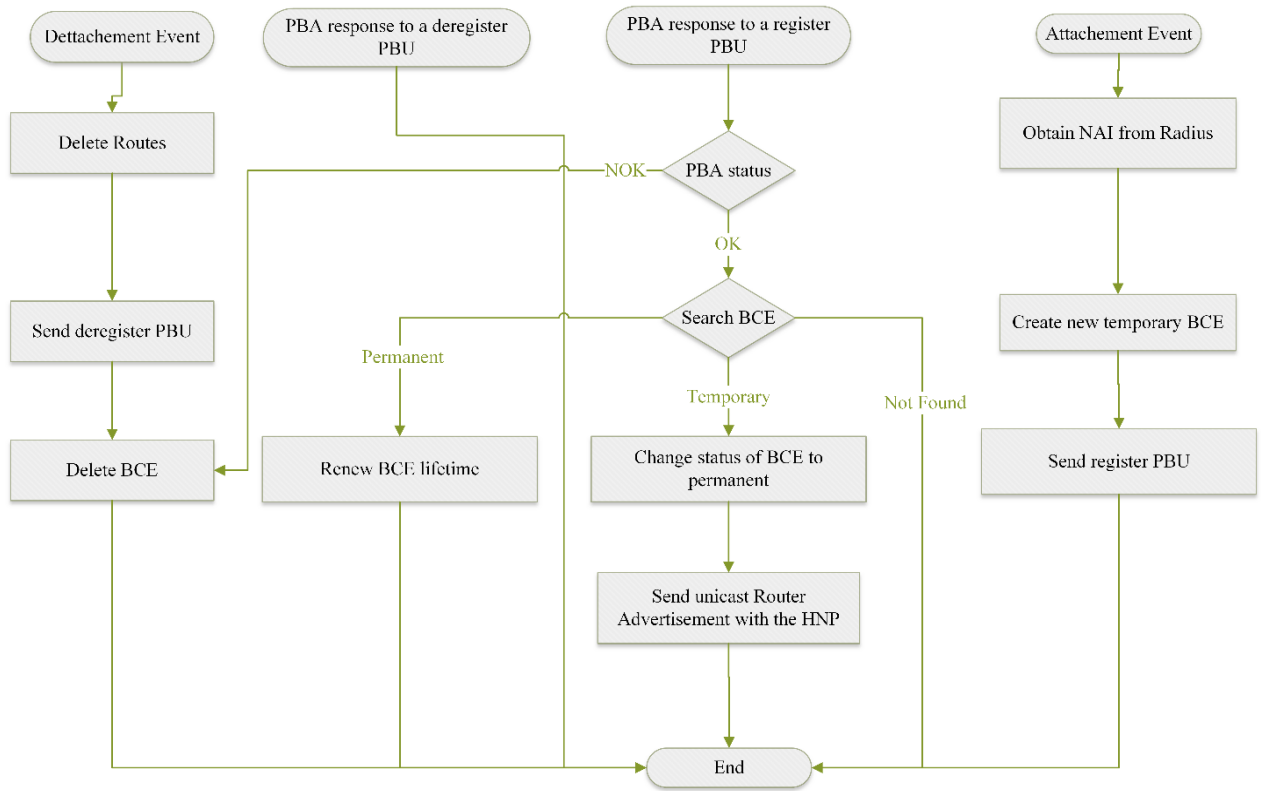

*Figure 5 - Finite State Machine of MAG*

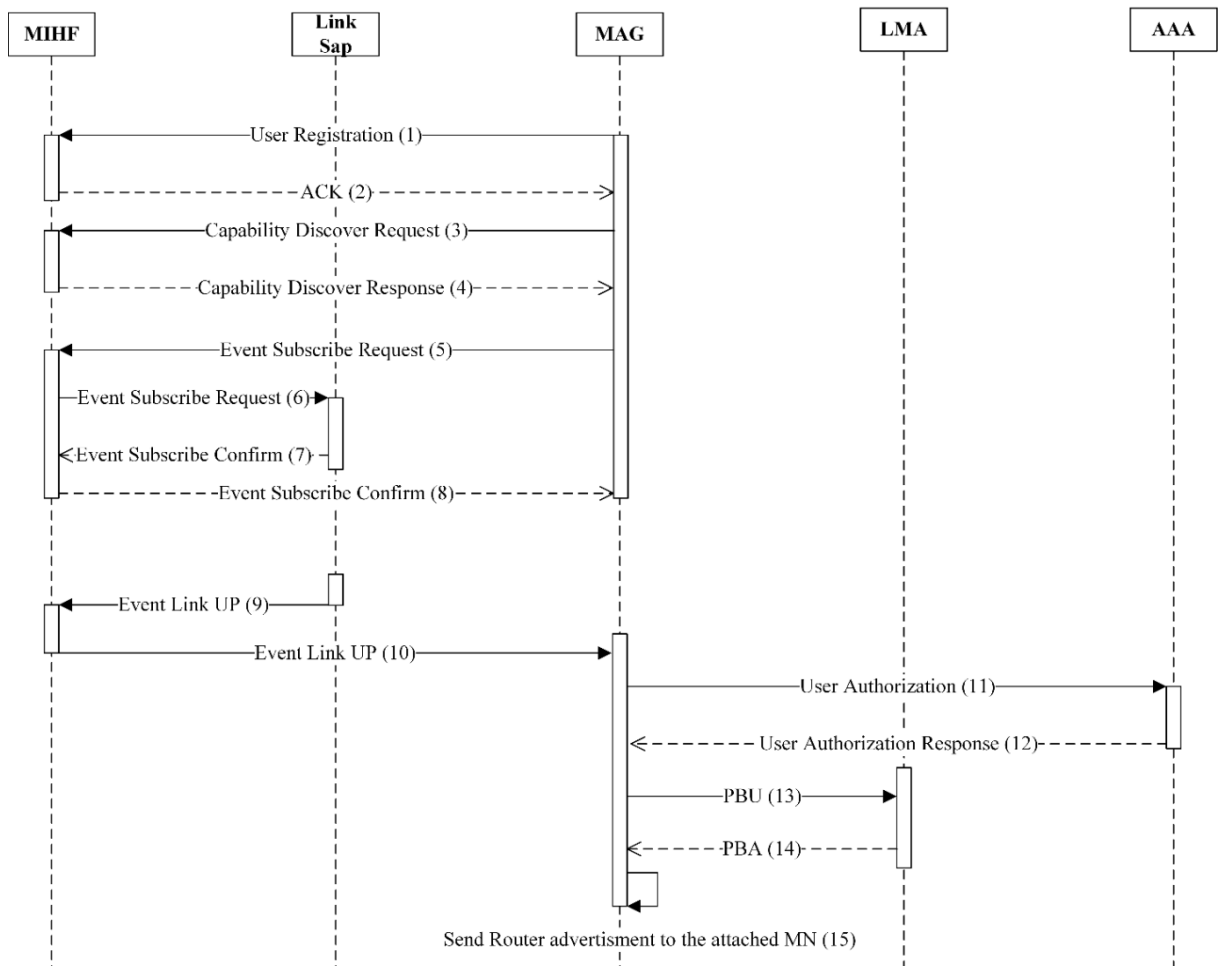

*Figure 6 - Sequence diagram representing the integrated operation of IEEE 802.21, AAA, and PMIPv6; it is also visualized how a mobile flow is tracked by our system*

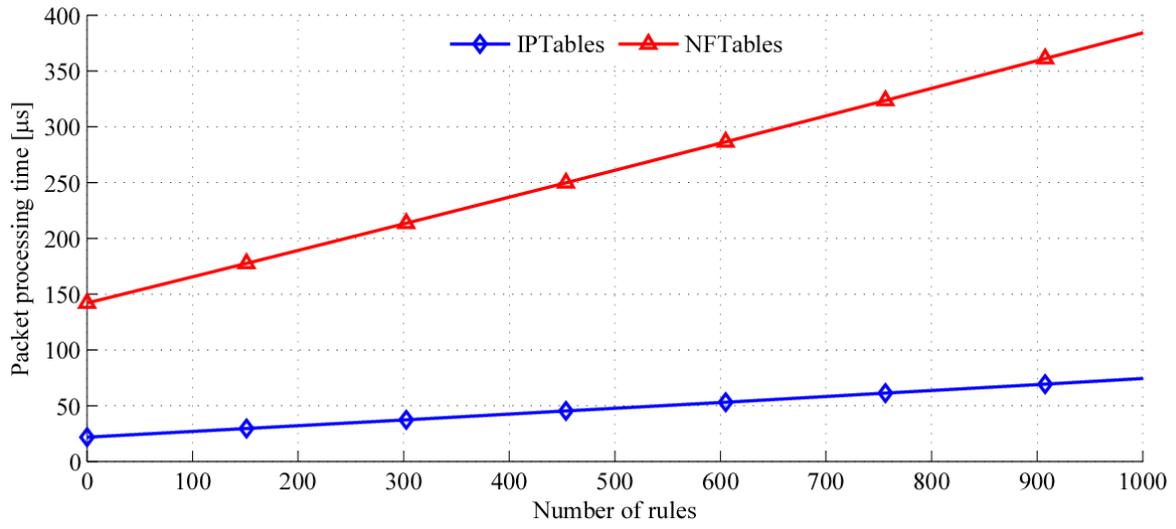

*Figure 7 – IPTables / NFTables packet processing time in relation to the number of rules*

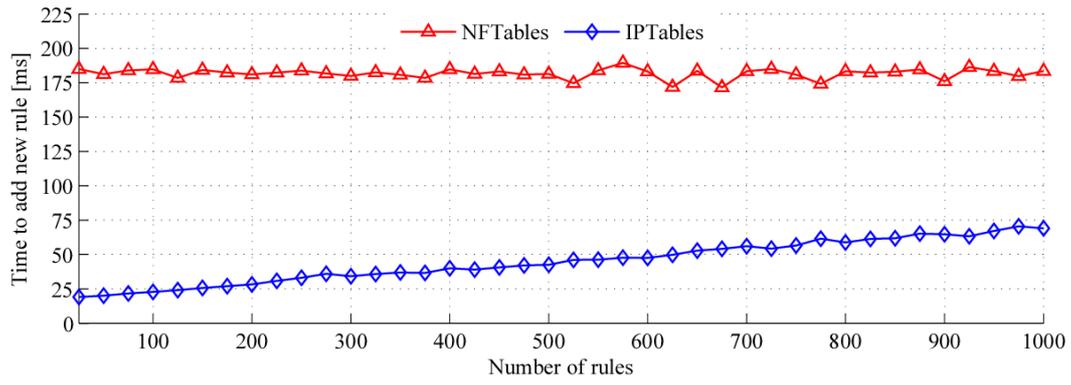

*Figure 8 – IPTables / NFTables insert time of a new rule in relation to the system's load*

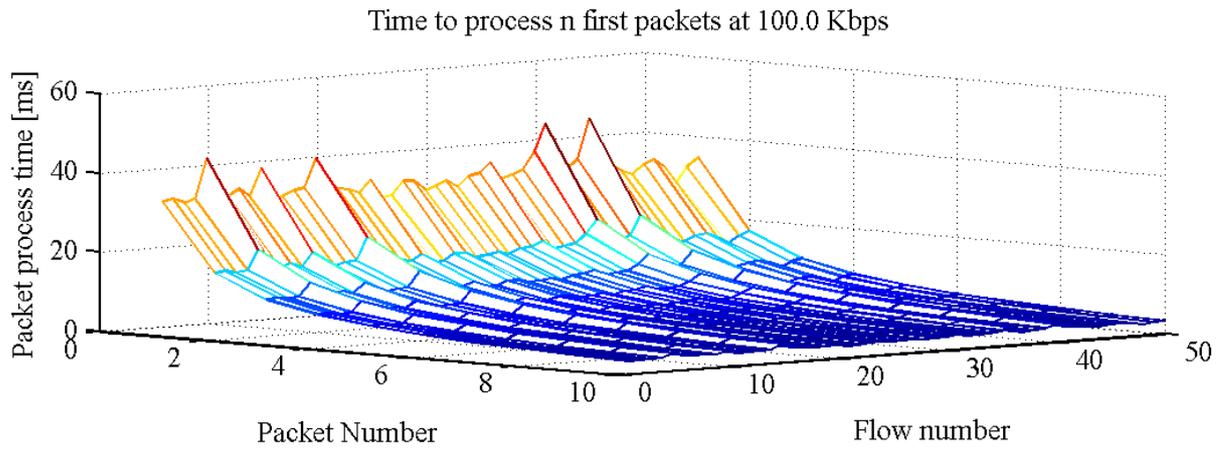

*Figure 9 – Packet processing time of the first set of n packets (packet number) of a new mobile flow, considering also the load imposed by a variable number of flows (flow number) already within the system*

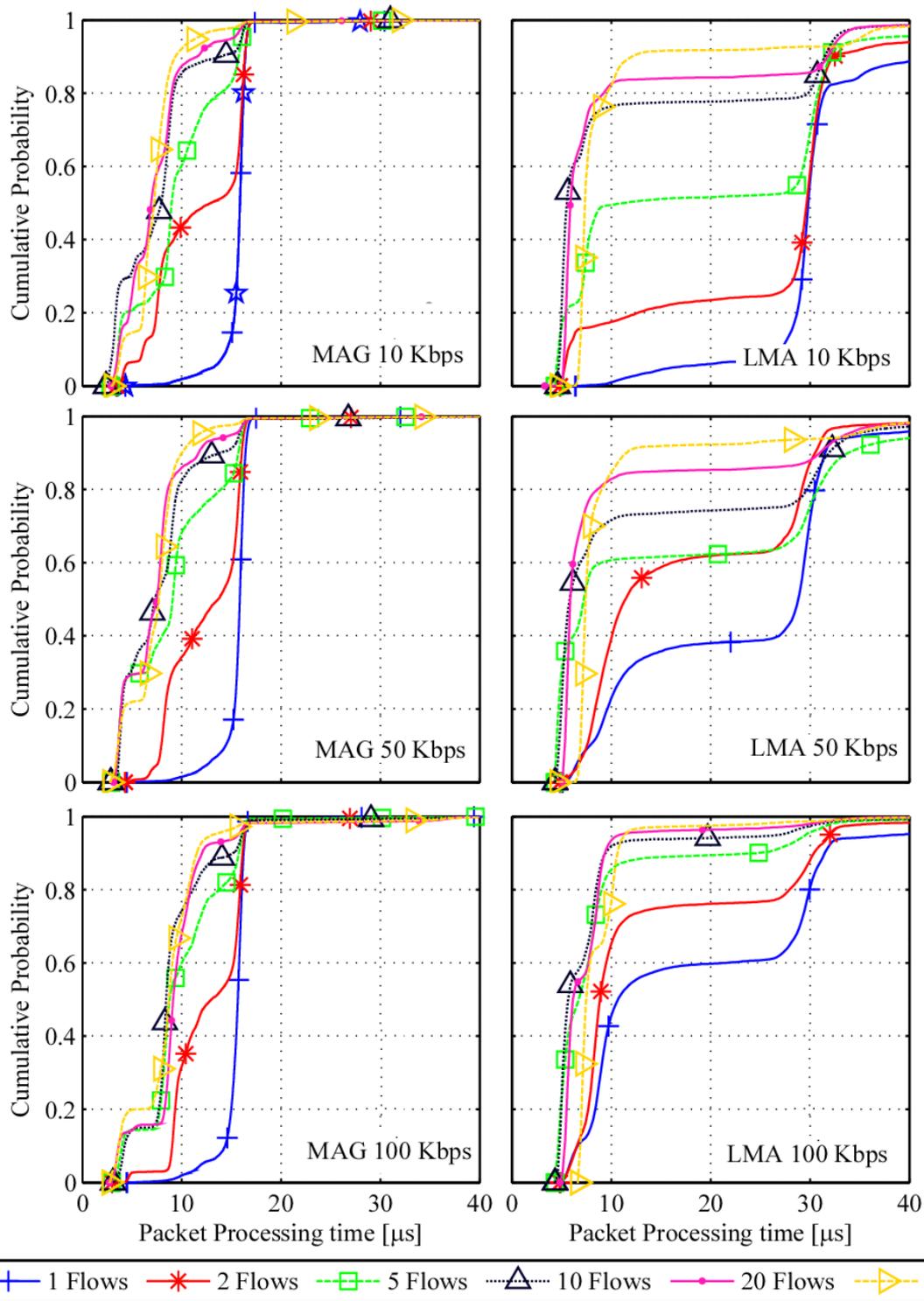

*Figure 10 – Statistics behaviour of MAG and LMA packet processing time for various flow throughputs*

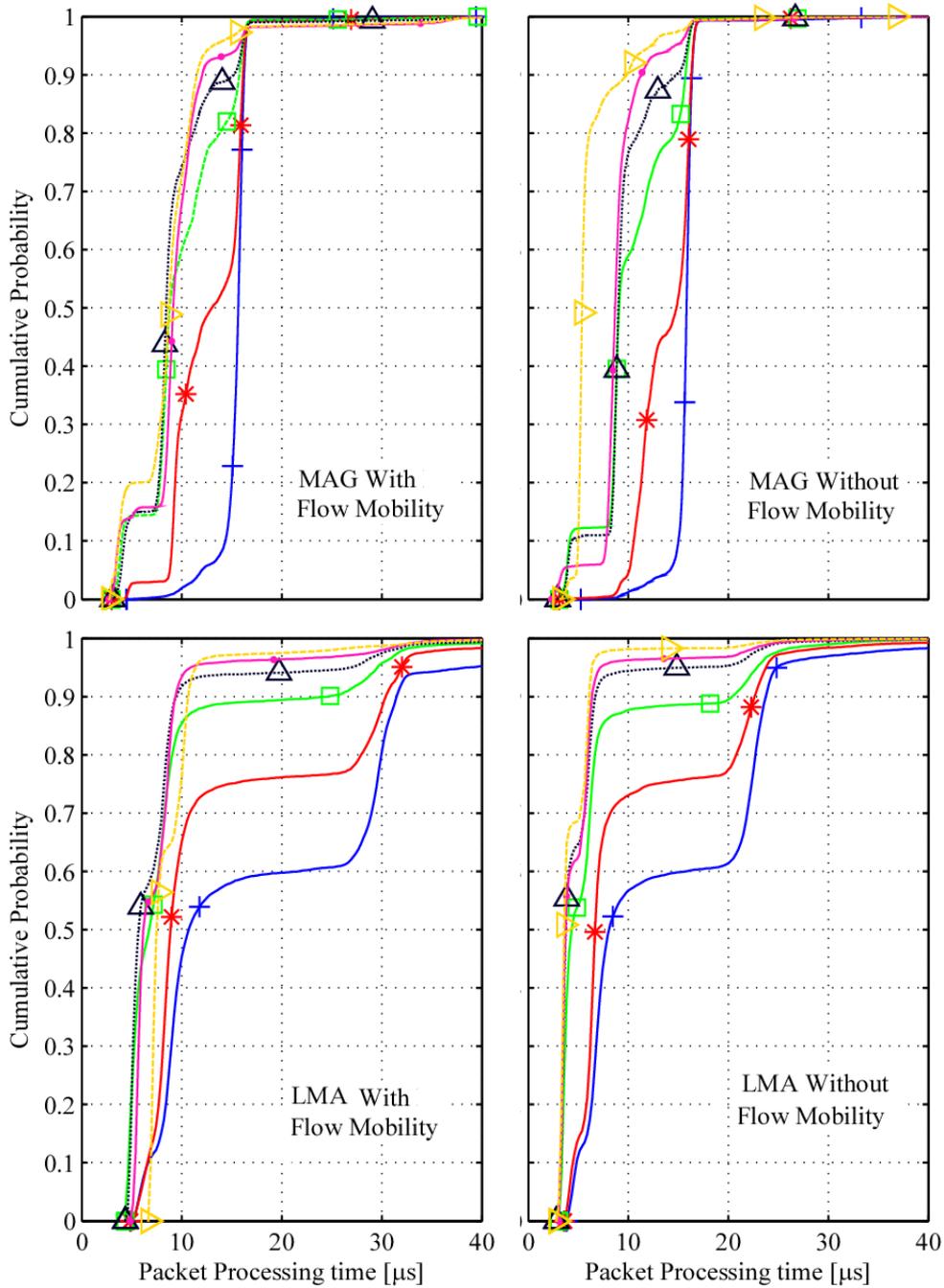
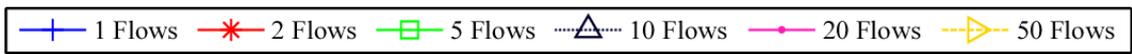

*Figure 11 – Statistics behaviour of LMA Packet processing time with and without flow mobility (flow throughput is always 100 Kbps)*

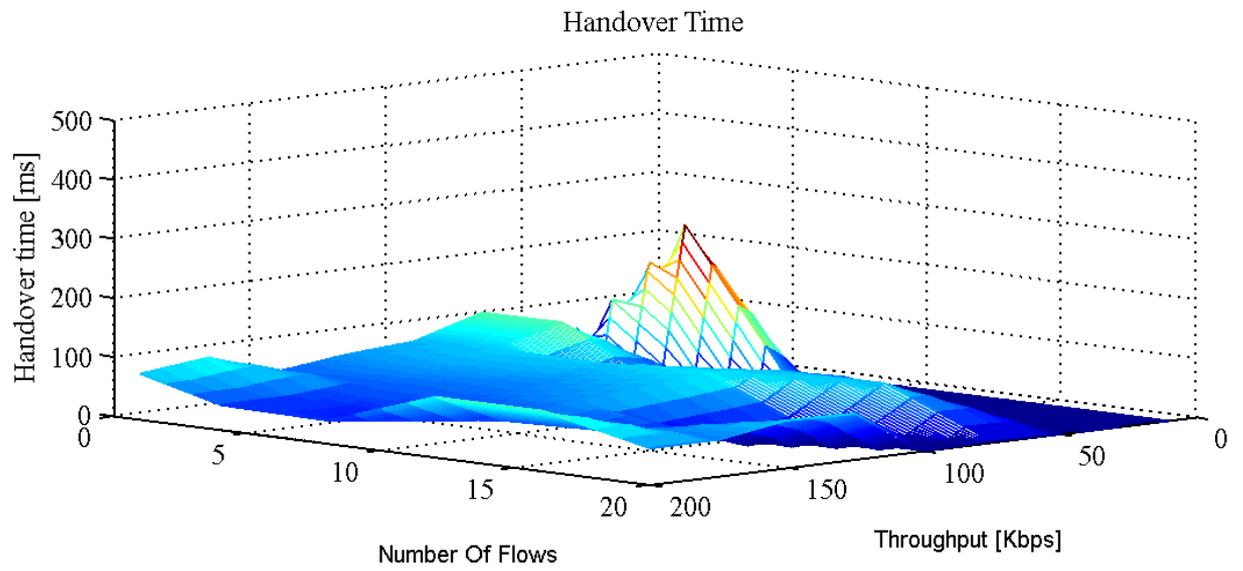

*Figure 12 - Handover Time trend vs. number of flows and flow throughput*